\documentclass[reprint,aps,pra,showpacs,superscriptaddress]{revtex4-1} % for arXiv
\usepackage{amsmath,amssymb}
\usepackage{mathrsfs}
\usepackage{color}
\usepackage{graphicx}% Include figure files
\usepackage{dcolumn}% Align table columns on decimal point
\usepackage{bm}% bold math
\usepackage{enumerate}
\usepackage{theorem}
\newtheorem{theorem}{Theorem}
\newtheorem{lemma}{Lemma}
\begin{document}
%\preprint{}
\title{Error probability analysis in quantum tomography: a tool for
evaluating experiments}



\author{Takanori Sugiyama}
\email{sugiyama@eve.phys.s.u-tokyo.ac.jp}
\affiliation{
Department of Physics, Graduate School of Science, The University of Tokyo,\\
7-3-1 Hongo, Bunkyo-ku, Tokyo, Japan 113-0033.
}
\author{Peter S. Turner}
\email{turner@phys.s.u-tokyo.ac.jp}
\affiliation{
Department of Physics, Graduate School of Science, The University of Tokyo,\\
7-3-1 Hongo, Bunkyo-ku, Tokyo, Japan 113-0033.
}
\author{Mio Murao}
\email{murao@phys.s.u-tokyo.ac.jp}
\affiliation{
Department of Physics, Graduate School of Science, The University of Tokyo,\\
7-3-1 Hongo, Bunkyo-ku, Tokyo, Japan 113-0033.
}
\affiliation{
Institute for Nano Quantum Information Electronics, The University of Tokyo,\\
4-6-1 Komaba, Meguro-ku, Tokyo , Japan 153-8505.
}
%\thanks{}
\date{\today}
\begin{abstract}
We expand the scope of the statistical notion of error probability, {\it i.e.}, how often large
deviations are observed in an experiment, in order to make it directly
applicable to quantum tomography.
We verify that the error probability can decrease at most exponentially in the number of
trials, derive the explicit rate that bounds this decrease, and
show that a maximum likelihood estimator achieves this bound.
We also show that the statistical notion of identifiability coincides with
the tomographic notion of informational completeness.  Our result
implies that two quantum tomographic apparatuses that have the same
risk function, (e.g. variance),
can have different error probability,
and we give an example in one qubit state tomography.  
Thus by combining these two approaches we can evaluate, in a reconstruction
independent way, the performance of such experiments more discerningly.
\end{abstract}
%\keywords{}
\maketitle
%\section{}
%\subsection{}
%\subsubsection{}
%
\section{Introduction}

   Many applications that make use of ``quantumness'' in order to
   outperform their classical counterparts have recently been proposed,
   especially in the field of quantum information.
   One of the main reasons for this increase has been the 
   dramatic development of experimental
   technologies, and many of the proposals have already given rise to
   experimentally realizable applications \cite{SchleichWalther07}.
   To confirm whether or not an apparatus constructed for an application works well,
   we need to compare its performance to a theoretical model.
   The standard method used for a thorough such comparison is called
   quantum tomography \cite{ParisRehacek04}.
   This paper is concerned mainly with the question of how to evaluate 
   measurement apparatuses used in quantum tomography.

   The theory of quantum tomography consists of experimental design
   methods and reconstruction schemes.
   Known parts of the experimental apparatus in a quantum tomographic
   experiment, (or at least those parts assumed to be known), are together called the \emph{tester}.
% Experimental design
   Experimental design methods are concerned with how good (or bad) the tester is
   for estimating the mathematical representation of the tomographic
   object ({\it e.g.} a quantum state, or a process).
%   A tester is called \emph{complete} if it makes it possible to identify an
%   arbitrary tomographic object from the experimental data
%   set of an infinite number of measurement trials.
%   The conditions required for the tester's completeness depend upon the
%   type of tomographic object and the experimental setup
%   \cite{Prugovecki77, D'Ariano03, Mohseni06}.
   Usually the goodness of the tester is evaluated by the error of the
   estimation result from its experimental data set.
%   Suppose that we can use two complete testers for a tomographic experiment.
%   If we could perform an infinite number of measurement trials, the
%   experimental results of two testers would lead to same estimation result
%   (the ``true'' tomographic object), and so the testers' performances are equivalent.
   In real experiments, we cannot perform an infinite
   number trials -- we need to estimate the true tomographic object from
   a finite number.
   This estimation procedure is called an estimator in statistical
   estimation theory and a reconstruction scheme in quantum
   tomography.
   The error of the estimation result depends upon the
   reconstruction scheme, and when evaluating a tester's
   performance, we usually focus upon the error in the case
   where the best reconstruction scheme is used.     

% Evaluation of statistical errors
% Risk functions
   Evaluating estimation errors on the reconstructed
   object is a problem of statistical estimation theory.
   There are two main approaches; one is to use a {\it risk function}
   and the other is to use {\it error probability}.
   We measure the difference between the true object and the estimate by a
   \emph{loss function}.
   A risk function is the average value of the loss function.  
   As the number of independent, identically distributed (iid) trials
   increases, it is known that the error, given by the risk function,
   of any unbiased estimator can decrease by at most
   the Cram\'{e}r-Rao bound, and a maximum likelihood estimator
   achieves the bound asymptotically \cite{Rao73}.
   The application of the Cram\'{e}r-Rao inequality to quantum
   tomography is studied in \cite{Gill00, Rehacek04, Bagan06, Nunn09}.
% Error probabilities
   On the other hand, an error probability is the probability that large deviations of the loss
   function are observed.
   It has been shown that the error probability can decrease at most 
   exponentially \cite{BGZ1980},
   and under some conditions, the bound is
   achieved (asymptotically) by a maximum likelihood estimator \cite{Shen01}.
   However, the explicit form of this bound
   has not been shown except for the case where the estimation setting can
   be reduced to one parameter estimation or the loss function is a
   Euclidean norm \cite{Bahadur1960, Bahadur1967, Hayashi02, HM00}  (in \cite{HM00} the applicability of the proof used for a Euclidean norm to more general loss functions is discussed). 
   In general, the estimated object has multiple parameters, 
   and the choice of the loss function depends upon the purpose of the experiment.
   A mean squared error can be unsuitable for some situations,
   especially those arising in quantum tomography.
   In order to be more useful in practice, the explicit form of the bound is
   needed in more generality.

   In this paper, by using Sanov's theorem \cite{Sanov57,DemboZeitouni98} from large deviation theory, 
   we derive the error probability inequality bounding general
   loss functions on a finite multiparameter space. 
   We prove that a maximum likelihood estimator achieves the
   equality under some conditions -- that are satisfied in quantum tomography --
   and give the explicit form of the lower bound.
   Our result indicates that two testers with same value of their risk functions
   can be different from an error probability viewpoint, which
   allows for more discerning comparisons of testers in quantum
   tomography.
   We also show that the required conditions for our inequality hold not only for tomography of quantum
   states, but also for that of quantum instruments \cite{D'Ariano01},
   which includes process and measurement tomography as specific cases.

   In section \ref{section:overview}, we overview the theory of quantum
   tomography using state tomography as an example.
   In section \ref{section:classical_estimation_theory}, we review
   classical statistical estimation theory, introducing the necessary aspects of error
   probability theory.
   In section \ref{section:main_result}, we give the main theorem and
   some analysis which includes an example -- the proof of the theorem is given in the Appendix.
   In section \ref{section:discussion}, we discuss some open problems,
   and conclude with a summary in section \ref{section:summary}.

\section{Overview of quantum tomography}\label{section:overview}

   Quantum tomography is classified by the tomographic object to be reconstructed:
   state tomography \cite{Smithey1993, Hradil1997, Banaszek1999} treats density
   operators, which describe states of quantum systems;
   process tomography \cite{Poyatos97, Chuang1997, Buzek1998,
   Fiurasek2001, Sacchi2001} treats linear,
   trace-preserving, and completely positive maps, which describe
   deterministic state transitions;
   POVM tomography \cite{Luis99, Fiurasek01_POVMT} treats positive
   operator-valued measures (POVMs), which are sets of
   positive-semidefinite operators describing the probability distributions
   obtained by measurements;
   instrument tomography \cite{D'Ariano01} treats quantum instruments,
   which are sets of linear, trace-decreasing, completely positive maps
   describing both probability distributions and state transitions
   caused by measurements.
   Here we briefly overview the theory of quantum state tomography, for
   simplicity.   

   The purpose of quantum state tomography is to identify the density
   operator characterizing the state of the quantum system of interest.   
   Let $\mathcal{H}$ and $\mathcal{S}(\mathcal{H})$ denote the Hilbert
   space corresponding to the system and the set of all density operators
   on $\mathcal{H}$, respectively.
   We assume that the dimension $d=\dim \mathcal{H}$ is finite.
   A density operator $\hat{\rho}$ on $\mathcal{H}$ can be linearly and
   bijectively parametrized by $d^2 -1$ independent real variables \cite{Kimura03,Byrd03}, 
   {\it i.e.}, $\hat{\rho} = \hat{\rho} (\bm{s})$, where $\bm{s}$ is in $\mathbf{R}^{d^2 -1}$.  
   Let us define the set of all parameters $S:=\{ \bm{s}\in
   \mathbf{R}^{d^2 -1}|\ \hat{\rho}(\bm{s})\in \mathcal{S}(\mathcal{H})\}$.
   Identifying the true density operator $\hat{\rho} \in
   \mathcal{S}(\mathcal{H})$ is equivalent to identifying the true
   parameter $\bm{s} \in S$. 
   Let $\bm{\Pi}=\{ \hat{\Pi}_{x} \}_{x\in \Omega}$ denote the POVM characterizing
   the tester used in the tomographic experiment %\cite{Note1},
   \footnote{In quantum tomography, it is possible to change the tester
  used in the next trial depending on the previous observation results. Such an
  experimental scheme is called {\protect \it adaptive}, and the rate of decrease in such a scheme is analyzed in \cite{HM00}. An experimental scheme
  which allows a global measurement on more than one system at once is called
  {\protect \it collective}. Both of these generalizations constitute
  significantly more complicated experiments that are not currently the norm,
  and are not treated in this paper.}
%   \footnote{In quantum tomography,
%     it is possible to change the tester used in the next trial depending on
%     the previous observation results. Such an experimental scheme is called
%   {\it adaptive}. 
%   An experimental scheme which allows a global measurement on 
%   more than one system at once is called {\it collective}.
%   Both of these generalizations constitute significantly more
%   complicated experiments that are not currently the norm, and are not
%   treated in this paper.}
   where $\Omega:=\{1,\ldots , M\}$.
   When the true density operator is $\hat{\rho}(\bm{s})$, the probability
   distribution $p_{\bm{s}}$ describing the tomographic experiment is given by 
   \begin{eqnarray}
      p_{\bm{s}}(x) = \mbox{Tr}[\hat{\rho}(\bm{s})\hat{\Pi}_{x}],\ x\in
       \Omega ,\label{eq:p=rhopi}
   \end{eqnarray} 
   where $\mbox{Tr}$ denotes the trace operation with respect to
   $\mathcal{H}$.
   (Note that in subsection \ref{subsection:DP_relation} a different trace
   operation, $\mbox{tr}$, is introduced.)
   Suppose that we perform $N$ measurement trials and obtain a
   data set $\bm{x}^N = (x_{1}, \ldots , x_{N})$, where $x_{i} \in \Omega$
   is the outcome observed in the $i$-th trial.
   Let $N_{x}$ denote the number of times that outcome $x$ occurs in $\bm{x}^N$, then
   $f_{N}(x):=N_{x}/N$ is the relative frequency of $x$ for the data set $\bm{x}^N$.
   In the limit of $N=\infty$, the relative frequency is equal to the
   true probability $p_{\bm{s}}(x)$.
   A tester is called \emph{informationally complete} if
   $\mbox{Tr}[\hat{\rho}(\bm{s}) \hat{\Pi}_{x})]=\mbox{Tr}[\hat{\rho} \hat{\Pi}_{x}]$ has a
   unique solution $\hat{\rho}$ for arbitrary $\hat{\rho}(\bm{s}) \in
   \mathcal{S}(\mathcal{H})$ \cite{Prugovecki77}.
   This condition is equivalent to that of the tester POVM $\bm{\Pi}$ being a
   basis for the set of all Hermitian matrices on $\mathcal{H}$.
   For finite $N$, the relative frequency and true probability are
   generally not the same, {\it i.e.}, there is unavoidable statistical
   error, and we need to choose an estimation procedure that takes the
   experimental result $\bm{x}^N$ to a density operator, {\it i.e.}, a
   reconstruction scheme.   

% Reconstruction schemes
   Reconstruction schemes are concerned with how best to derive the mathematical
   representation of the tomographic object from the obtained experimental data, and are called
   \emph{estimators} in statistical estimation theory, where the analysis of the estimation
   precision (or estimation error) is very important.
   In actual experiments, there are two sources of imprecision: statistical
   errors and systematic errors.
   As mentioned above, statistical error is caused by the finiteness of the total number
   of measurement trials, and is unavoidable in principle.
   Systematic error is caused by our lack of knowledge about the
   tester, that is, the difference between the true tester and what we
   believe to be the true tester.    
   Usually, the effect of the systematic error is approximated by
   introducing a model, and is assumed to be known.
   Therefore, the analysis of the estimation error is usually reduced to that of
   the statistical error. 
   To date at least five reconstruction schemes have been proposed,
   namely linear \cite{Smithey1993, Poyatos97, Chuang1997, Luis99, D'Ariano01}, maximum likelihood
   \cite{Hradil1997, Fiurasek2001, Fiurasek01_POVMT}, Bayesian
   \cite{Buzek98, Schack01, Fuchs04}, maximum entropy \cite{Buzek00}, and
   norm minimization \cite{Kosut08}.
   The effect of statistical errors on the reconstructed object depends
   upon the scheme used, hence the main problem is how to quantify
   the effect of the statistical error on the reconstructed object, and how
   to do so as rigorously as possible. 

   It is natural to consider a linear reconstruction scheme, which demands that we find a $d \times d$
   matrix $\hat{\rho}^{\mathrm{l}}$ satisfying 
   \begin{eqnarray}
      \mbox{Tr}[\hat{\rho}^{\mathrm{l}}\hat{\Pi}_{x}] = f_{N}(x),\ x\in \Omega . \label{eq:Linear}
   \end{eqnarray} 
   However, Eq.(\ref{eq:Linear}) does not always have a solution,
   and even when it does, although the solution is
   Hermitian and normalized, it is not guaranteed that $\hat{\rho}^{\mathrm{l}}$ is positive semidefinite. 
   A maximum likelihood reconstruction scheme addresses these problems.
   The estimated matrix $\hat{\rho}^{\mathrm{ml}}$ is defined as
   \begin{eqnarray}
      \hat{\rho}^{\mathrm{ml}} := \mbox{argmax}_{\hat{\rho} \in \mathcal{S}(\mathcal{H})}
                    \prod_{i=1}^{N}\mbox{Tr}[\hat{\rho} \hat{\Pi}_{x_i}].
   \end{eqnarray}
   It can be shown that when $\hat{\rho}^{\mathrm{l}} \in \mathcal{S}(\mathcal{H})$,
   $\hat{\rho}^{\mathrm{l}}=\hat{\rho}^{\mathrm{ml}}$ holds. 
   We will concern ourselves with maximum likelihood reconstructions
   here, as we will see that they are optimal in the sense that they can saturate the bounds we are considering.

\section{Overview of classical statistical estimation}\label{section:classical_estimation_theory}

   In this section, we introduce the notation and terminology of
   classical statistical estimation theory that we use to arrive
   at our main results. We also review the necessary aspects of
   error probability.
   For the reader familiar with quantum Fisher matrices, we justify
   our use of classical estimation theory, or the classical Fisher matrix,
   in subsection \ref{subsection:quantum_estimation_theory}.

   Let $(\Omega , \mathscr{B}, P)$ be a probability space, 
   denoting a sample space, a Borel algebra of subsets of
   the sample space, and a measure that assigns probabilities to those subsets,
   respectively (the Borel structure simply assures that combinations of
   subsets of events get assigned probabilities in a sensible fashion,
   {\it e.g.} $P(Y\cup Z)=P(Y)+P(Z)$ for disjoint $Y$ and $Z$).
   Define the $N$-fold direct product $\Omega^{N}:=\Omega \times
   \cdots \times \Omega$ as the space of sequences of events.  
   Let $\bm{x}^N = \{ x_{1}, \cdots , x_{N}\},\ x_i \in \Omega ,$ be a
   sequence of iid observations of the sample space $\Omega$.
   We assume that the sample space is finite (see subsection
   \ref{subsection:infinite_sample_space} for a discussion of infinite
   spaces).
   Suppose that the probability space admits a statistical
   model $\mathcal{P}_{\Theta}=\{P_{\theta}; \theta \in \Theta\}$ 
   that assigns a valid probability measure to each parameter $\theta$
   in $\Theta$ which is an subset of the $k$-dimensional Euclidean space
   $\mathbf{R}^k$, the closure $\bar{\Theta}$ is compact, and the
   interior $\Theta^{o}$ is open.
   The quantum state parameter space $S$ from the last section is an
   example of such a $\Theta$, where the statistical model is given by
   Eq.(\ref{eq:p=rhopi}).
   We assume that each measure $P_{\theta}$ has a
   probability distribution $\{ p_{\theta}(x) \}_{x \in \Omega}$ satisfying
   $P_{\theta}(Y)=\sum_{x \in Y}p_{\theta}(x)$ where $Y\in\mathscr{B}$.
   A probability measure $P_{\theta}$ and the probability distribution
   $p_{\theta}$ have a one-to-one correspondence for any $\theta \in \Theta$, and we do not
   distinguish between $\mathcal{P}_{\Theta}$ and $\{p_{\theta}; \theta
   \in \Theta \}$.
   Let $\mathcal{P}(\Omega)$ denote the set of all probability
   distributions with the sample space $\Omega$, then
   $\mathcal{P}_{\Theta}\subseteq \mathcal{P}(\Omega)$.
   Let $P_{\theta}^{(N)}$ denote the $N$-fold product probability
   measure $P_{\theta} \times \cdots \times P_{\theta}$.  

% Estimators and estimates
    Let $g$ denote a map from the parameter space $\Theta$ to a metric
    space $\Gamma$.  An estimator of $g(\theta)$ is a set
    of maps $\varphi = \{\varphi_{1}, \varphi_{2}, \ldots \}$, (one for
    each number of trials $N$), from observation results $\bm{x}^N$ to
    $\Gamma$.
    Each $\varphi_{N}(\bm{x}^{N})$ is called the estimate of $\bm{x}^{N}$.
% ML estimator
    A maximum likelihood estimator $\theta^{\mathrm{ml}} = \{\theta^{\mathrm{ml}}_{1},
    \theta^{\mathrm{ml}}_{2}, \ldots \}$ of $\theta$ is defined as 
    \begin{eqnarray}
       \theta^{\mathrm{ml}}_{N}(\bm{x}^{N})= \mbox{argsup}_{\theta \in \Theta}P_{\theta}^{(N)}(\{\bm{x}^{N}\}).
    \end{eqnarray}
%    An induced maximum likelihood estimator $\varphi^{\mathrm{iml}} = \{\varphi^{\mathrm{iml}}_{1},
%    \varphi^{\mathrm{iml}}_{2}, \ldots \}$ is defined as 
%    \begin{eqnarray}
%       \varphi^{\mathrm{iml}}_{N}(\bm{x}^{N})= \mbox{argmax}_{\gamma \in
%	\Gamma} \mbox{argmax}_{\theta \in \Theta ; g(\theta) = \gamma}P_{\theta}^{(N)}(\{\bm{x}^{N}\}).
%    \end{eqnarray}
    
% Loss functions
    A map $D$ from $\Gamma \times \Gamma$ to $\mathbf{R}$ is called a loss
    function on $\Gamma$ when $D$ satisfies the following two conditions: 
    (i) $\forall a,b \in \Gamma$, $D(a,b)\ge 0$,
    (ii) $\forall a \in \Gamma$, $D(a,a)=0$. 
    We introduce three additional conditions:
    (iii) $\forall a,b \in \Gamma$, $D(a,b) = D(b,a)$,
    (iv) $\forall a,b,c \in \Gamma$, $D(a,b)\le D(a,c)+D(c,b)$,
    (v) $\forall a,b \in \Gamma$, $D(a,b)= 0 \,\Rightarrow\, a=b$.
    A loss function satisfying conditions (iii) and (iv) is called a semi-distance, and
    a semi-distance satisfying condition (v) is called a distance.
    For example, let us define a function $g$ from $\Theta$ to $\mathbf{R}$
    as $g(\theta)=\|\theta \|, \theta\in\Theta$, where $\|\cdot \|$ is the
    Euclidean norm on $\mathbf{R}^k$.
    Then $|g(\theta)-g(\theta^{\prime})|$ is a semi-distance on $\Theta$
    ($\theta, \theta^{\prime}\in \Theta$) and $|a-b|$ is a distance on
    $\mathbf{R}$ ($a,b\in\mathbf{R}$). 
    In general, a loss function is not necessarily a distance.
    A loss function satisfying condition (v) is called a pseudo-distance %\cite{Note2}.
    \footnote{The terminology differs by textbook.}.
    The Kullback-Leibler divergence introduced below is
    an example of pseudo-distance that is not also a distance.
    If a loss function $D$ on $\mathbf{R}^k$ is sufficiently smooth
    and it can be approximated by the Hesse matrix $H_{a}$ up to second order,
    then $H_{a}$ is positive semidefinite for all $a \in \mathbf{R}^k$ from
    condition (i), and if the loss function $D$ is a pseudo-distance,
    then $H_{a}$ is positive definite for all $a \in \mathbf{R}^k$.

% Evaluation methods for estimation errors
    There are at least two methods to evaluate an estimation error by
    using loss function.
% Risk functions
    One is a method using risk functions.
    An $N$-trial risk function $\bar{D}^{(N)}$ is defined as the expectation value of the loss
    function between an estimate and the true object, given by $\theta$;
    \begin{eqnarray}
       \bar{D}^{(N)}:=E_{\theta}^{(N)}[D(\varphi_{N}(\bm{x}^N),g(\theta))],
    \end{eqnarray}  
    where $E_{\theta}^{(N)}[f(\bm{x}^N)]=\sum_{\bm{x}^N \in\Omega^N} p_{\theta}(\bm{x}^N )f(\bm{x}^N)$
    is the expectation value of a function $f$ on $\Omega^N$. 
    When $\Gamma=\mathbf{R}^l$, for any unbiased estimator ($\varphi$
    satisfying $E_{\theta}^{(N)}[\varphi_{N}(\bm{x}^N)]=g(\theta)$ for any $N$
    and $\theta \in \Theta$), the Cram\'{e}r-Rao inequality 
    \begin{eqnarray}
       E_{\theta}^{(N)}[(\varphi_{N}(\bm{x}^N)-g(\theta))(\varphi_{N}(\bm{x}^N)-g(\theta))^{T}]
       \notag \\
      \ge \frac{1}{N} \frac{\partial g}{\partial \theta}^{T}F_{\theta}^{-1}\frac{\partial g}{\partial \theta}
    \end{eqnarray} 
    holds under some regularity conditions, where
    $\bigl(\frac{\partial g}{\partial \theta}\bigr)_{\alpha
    \beta}:=\frac{\partial g_{\beta}}{\partial
    \theta_{\alpha}}\ (\alpha = 1, \ldots , k; \beta = 1, \ldots , l)$ is the
    Jacobian and
    $F_{\theta}^{-1}$ is the Moore-Penrose generalized inverse of the Fisher matrix\\
    $F_{\theta}:= \sum_{x \in \Omega}p_{\theta}(x) [\nabla_{\theta}\log
    p_{\theta}(x)][\nabla_{\theta}\log p_{\theta}(x)]^{T}$.
    Asymptotically a maximum likelihood estimator achieves the equality
    under some conditions \cite{Rao73}.

% Error probability
    The other is a method using error probabilities.
    We call
    \begin{eqnarray}
       P^{(N)}_{\epsilon}(\theta)
       :&=& P_{\theta}^{(N)}(D(\varphi_N (\bm{x}^N), g(\theta))>\epsilon )\\
        =&P_{\theta}^{(N)}&(\{\bm{x}^N \in
	\Omega^N ; D(\varphi_N (\bm{x}^N), g(\theta))>\epsilon \})
    \end{eqnarray}
    an error probability with a threshold $\epsilon >0$.
    An estimator $\varphi$ is called {\it $($weakly$)$ consistent} in the
    loss function $D$ if 
    \begin{eqnarray}
       P^{(N)}_{\epsilon}(\theta) \to 0 \ \mbox{as} \ N \to \infty
    \end{eqnarray}
    holds for any $\epsilon >0$.
% Identifiability condition
    The conditions under which a maximum likelihood estimator is consistent includes the {\it identifiability condition} \cite{Wald49} on a statistical model $\mathcal{P}_{\Theta}$:
    for any $\theta \in \Theta^{o}$ and $\theta^{\prime} \in \Theta$, if $\theta\neq
    \theta^{\prime}$, then there exists at least one outcome
    $x \in \Omega$ satisfying $p_{\theta}(x) \neq
    p_{\theta^{\prime}}(x)$ \cite{Bahadur1960,Bahadur1967}. 
    Let us define 
    \begin{eqnarray}
       R_{\epsilon}(\theta):=\inf_{\theta^{\prime}\in \Theta}\{ K(p_{\theta^{\prime}}\| p_{\theta}); D(g(\theta^{\prime}), g(\theta))>\epsilon \},\label{eq:Rdef}
    \end{eqnarray}
    where $K(q\| p)=\sum_{x \in \Omega}q(x)\log \frac{q(x)}{p(x)}$ is
    called the Kullback-Leibler divergence (also known as the relative entropy).
    When $g$ is injective and $D$ is a distance, for any weakly consistent estimator in $D$,
    \begin{eqnarray}
       \varliminf_{N \to \infty}\frac{1}{N}
      \log P^{(N)}_{\epsilon}(\theta) \ge - R_{\epsilon}(\theta)\label{eq:BGZ}
    \end{eqnarray}
    holds \cite{BGZ1980}.
    It is known that in general the lower bound of Eq.(\ref{eq:BGZ}) is not
    attainable by any estimate \cite{Rukhin83}. 
    If we consider the limit $\epsilon \to 0$, under some conditions
    (including the identifiability condition), a maximum likelihood
    estimator achieves the equality, that is,
    \begin{eqnarray}
       \lim_{\epsilon \to 0}\lim_{N \to \infty}\frac{1}{\epsilon^{u}N}
      \log P^{(N)}_{\epsilon}(\theta) = - r(\theta),
    \end{eqnarray} 
    where $u$ is a real number suitable for $D$ and
    $r(\theta):=\lim_{\epsilon \to 0}\frac{R_{\epsilon}(\theta)}{\epsilon^u}$.
    The explicit forms of the rate are known for two specific cases.
    The first is the case where $\Gamma=\mathbf{R}$ and $D$ is the absolute
    value. In this case, the order $u$ is 2 and the explicit form of the lower bound is
    known to be \cite{Bahadur1960, Bahadur1967} 
    \begin{eqnarray}
    r(\theta)=\frac{1}{2\nabla_{\theta}g(\theta)\cdot
    F_{\theta}^{-1}\nabla_{\theta}g(\theta)}\label{eq:r1}.
    \end{eqnarray}
    The second is the case where $\Gamma=\mathbf{R}^{k}$, $D$ is the
    Euclidean distance on $\mathbf{R}^{k}$, and the order $u$ is again 2;
    the explicit form is \cite{HM00}
    \begin{align}
       r(\theta) = \frac{1}{2}\inf_{\bm{a}\in\mathbf{R}^{k}; \|\bm{a}
     \|=1} \bm{a} \cdot F_{\theta} \bm{a}.\label{eq:r2}
    \end{align}
    For more general $\Gamma$ or $D$, however, the explicit form of the lower bound is not known.
    Quantum state tomography corresponds to the case where
    $\Gamma = \mathbf{R}^{d^2 -1}$, and the standard loss function is
    the square of the fidelity distance 
    $D_{F}(\hat{\rho} , \hat{\rho}^{\prime}):= 1-
    \mbox{Tr}[\sqrt{\sqrt{\hat{\rho}}\hat{\rho}^{\prime}
    \sqrt{\hat{\rho}}}]^2$ or the square of the trace distance $D_{T}(\hat{\rho} , \hat{\rho}^{\prime}):=\mbox{Tr}[|\hat{\rho} -\hat{\rho}^{\prime}|]^2$.  
    In this paper, we extend the above results to multiparameter spaces
    and more general loss functions such as these that are directly applicable to quantum
    tomography, and give the explicit form of the lower bound.
    We apply our result to one qubit state tomography and show that it makes it possible to evaluate the performance of an experimental apparatus in greater detail.
    We also give quantum tomography conditions equivalent to the identifiability condition in classical estimation theory.  

\section{Main Result and analysis}\label{section:main_result}

   \subsection{Main theorem}\label{subsection:main_theorem}

   For simplicity we consider quantum state tomography.
   Suppose that we use a loss function $D$ on $\mathcal{S}(\mathcal{H})$.
   Let us define a loss function $\Delta$ on $S$ as 
   $\Delta (\bm{s}, \bm{s}^{\prime}):= D(\hat{\rho} (\bm{s}), \hat{\rho}
   (\bm{s}^{\prime}))$ $\forall \bm{s}, \bm{s}^{\prime} \in S$.
   Assume that $\Delta$ is sufficiently smooth.
   Let $S^{o}$ denote the interior of $S$.
   We define a same point Hesse matrix $H_{\bm{s}}$ for a two variable function $f$ on
   $S\times S$ as $\nabla_{\bm{s}^{\prime}}\nabla_{\bm{s}^{\prime}}f(\bm{s}^{\prime}, \bm{s})|_{\bm{s}^{\prime}=\bm{s}} = [ \partial_{s^{\prime\alpha}}\partial_{s^{\prime\beta}} f(\bm{s}^{\prime},\bm{s})|_{\bm{s}^{\prime}=\bm{s}}]$.
   In the following theorem, we assume the second order approximatability on the loss function.
   We choose the order of the error probability threshold $\epsilon$ to be two, in agreement with the Euclidean distance.   

   \begin{theorem}\label{theorem:main}
      Suppose that $\Delta$ is a pseudo-distance on $S$ with a non-zero same point
    Hesse matrix $H_{\bm{s}}$.
   If $\bm{s} \in S^{o}$, for an arbitrary consistent estimator $\bm{s}^{\mathrm{est}}$,
   the following inequality holds:
   \begin{eqnarray}
      \varliminf_{\epsilon \to 0}\varliminf_{N\to \infty}\frac{1}{\epsilon^2 N}\log
   P^{(N)}_{\bm{s}}(\Delta(\bm{s}^{\mathrm{est}}_{N}, \bm{s})>
   \epsilon^2 ) \notag \\
   \ge -1/\sigma_{1}( \sqrt{H_{\bm{s}}}F_{\bm{s}}^{-1}\sqrt{H_{\bm{s}}}),\label{eq:main1}
% -\inf_{d(\theta , \theta)\ge \epsilon}\{ K(\theta | \theta) \}.
   \end{eqnarray}
   where $\sigma_{1}(A)$ is the maximal eigenvalue of an Hermitian matrix $A$.
   Furthermore, when the tester is informationally complete, a maximum
    likelihood estimator $\bm{s}^{\mathrm{ml}}$ is consistent and
    achieves the equality in Eq.(\ref{eq:main1}),{\it i.e.}, 
   \begin{eqnarray}
      \lim_{\epsilon \to 0}\lim_{N\to \infty}\frac{1}{\epsilon^2 N}\log
   P^{(N)}_{\bm{s}}(\Delta(\bm{s}^{\mathrm{ml}}_{N}, \bm{s})> \epsilon^2 ) \notag \\
   = -1/\sigma_{1}( \sqrt{H_{\bm{s}}}F_{\bm{s}}^{-1}\sqrt{H_{\bm{s}}})\label{eq:main2}
% -\inf_{d(\theta , \theta)\ge \epsilon}\{ K(\theta | \theta) \}.
   \end{eqnarray}
   holds.
   \end{theorem}

   The detailed proof of Theorem \ref{theorem:main} appears in the Appendix
   -- here we give an outline.
   The proof is divided into six parts.
   For parts one through five, we do not assume that the probability
   distributions are quantum mechanical; we only assume that they are
   sufficiently differentiable and that the parameter space is compact.
   Only in the sixth part does quantum mechanics arise.
% 1  
   In Lemma 1, by using the same logic as the proof of Eq.(\ref{eq:BGZ}) in
   \cite{BGZ1980}, we show that Eq.(\ref{eq:BGZ}) holds for any
   estimator consistent not only in distances but also in pseudo-distances.
% 2
   Lemma 2 is introduced in order to calculate the infimum in Eq.(\ref{eq:Rdef}) directly.
% 3
   We use this in Lemma 3, where we obtain the explicit form of the bound on the rate, and obtain Eq.(\ref{eq:main1}).
% Emrirical measures and Sanov's theorem
   Next we introduce Sanov's theorem, a large deviation theorem that,
   roughly speaking, gives the rate of the probability of observing a relative
   frequency that differs from the true probability distribution.
% 4
   Lemma 4 uses the compactness of the parameter space and Sanov's
   theorem to prove that the error probability of a maximum likelihood
   estimator decreases exponentially if the identifiability condition is satisfied.
   Then, the maximum likelihood estimator is consistent and satisfies Eq.(\ref{eq:main1}). 
% 5
   In Lemma 5, we calculate the rate of decrease of the maximum likelihood
   estimator directly by using Sanov's theorem and Lemma 3,
   and show that the rate coincides with the lower bound in Eq.(\ref{eq:main1}).
   Hence, we obtain Eq.(\ref{eq:main2}), subject to the identifiability condition.
% 6
   Finally, we prove that in quantum state tomography the identifiability condition is equivalent to
   the informational completeness of the tester, which we present as Lemma 6.
   Together these lemmas prove Theorem \ref{theorem:main}.

   Note that in the proof we assume the compactness of the parameter
   space (in Lemmas 1 to 5) and the linear parametrizability of
   probability distributions (in Lemma 6).
   These assumptions hold for any quantum operator.
   Also, the concept of identifiability applies to the tomographic
   completeness of states equally well as it does to the informational
   completeness of measurements, which can be shown using the same logic
   as that of Lemma 6.
   Thus theorem \ref{theorem:main} holds for all types of quantum tomography.
   The dimension of the parameter space $k$ depends upon the type of
   quantum tomography: $k=d^2 -1$ and $d^4 -d^2$ for state and process
   tomography, respectively.
   For POVM and instrument tomography, $k=(M-1)d^2$ and $Md^4 -d^2$ respectively,
   where $M$ denotes the number of measurement outcomes.   
  
   \subsection{Meaning of the lower bound}

   Theorem \ref{theorem:main} indicates that in quantum tomography, if we have
   a sufficiently large data set, the error probability of any consistent
   estimator with a small threshold can decrease at most exponentially, and the
   rate is bounded by an estimator-independent function
   $1/\sigma_{1}( \sqrt{H_{\bm{s}}}F_{\bm{s}}^{-1}\sqrt{H_{\bm{s}}})$.
   Also, the bound is achievable by a maximum likelihood estimator. 
   Therefore, from the error probability viewpoint, if we can perform a large
   number of measurement trials, a maximum likelihood reconstruction scheme is optimal. 
   We can evaluate the performance of a given tester by the size of the
   maximal eigenvalue of the matrix
   \begin{eqnarray}
      G_{\bm{s}}:= \sqrt{H_{\bm{s}}}F_{\bm{s}}^{-1}\sqrt{H_{\bm{s}}}.
   \end{eqnarray}
   Testers with smaller maximal eigenvalues are better. 
   The inverse Fisher matrix $F_{\bm{s}}^{-1}$ alone characterizes the
   parameter-identifiability of the tester with respect to the Euclidean
   distance because the Hesse matix of the square of the Euclidean distance
   $\Delta^{E}(\bm{s},\bm{s}^{\prime}):=\|\bm{s}-\bm{s}^{\prime} \|^2$
   is $2I$, and we obtain 
   \begin{eqnarray}
      \frac{1}{\sigma_{1}(G_{\bm{s}})}
     &=& \frac{1}{2\sigma_{1}(F_{\bm{s}}^{-1})}\\
     &=& \frac{1}{2}\sigma_{k}(F_{\bm{s}})\\
     &=& \frac{1}{2}\inf_{\bm{a}\in \mathbf{R}^k
      ;\|\bm{a}\|=1}\bm{a}\cdot F_{\bm{s}}\bm{a},
   \end{eqnarray}
   where $\sigma_{k}(A)$ is the minimal eigenvalue of an Hermitian
   matrix $A$.
   This result coincides with the known result of Eq.(\ref{eq:r2}).
   The loss function $\Delta$ characterizes the purpose of the estimation
   (what we want to know), and the same point Hesse matrix $H_{\bm{s}}$
   modifies the inverse Fisher matrix from the Euclidean distance to
   the loss function $\Delta$ on $S$.
   Therefore the matrix $G_{\bm{s}}$ characterizes the parameter-identifiability of
   the tester with a modification according to our estimation purpose.

   \subsection{Relation to risk functions}\label{subsection:DP_relation}

   If we assume sufficient smoothness of a loss function $\Delta$ on $S$
   and informational completeness on the tester, a generalized
   Cram\'{e}r-Rao inequality can be derived, {\it i.e.}, for any
   unbiased estimator, the following inequality holds:
   \begin{eqnarray}
      \bar{\Delta}^{(N)}\ge \frac{\mbox{tr}[H_{\bm{s}}F_{\bm{s}}^{-1}]}{2N}
    +o(\frac{1}{N}), \label{eq:gCRineq}
   \end{eqnarray}
   where $\mbox{tr}$ denotes the trace operation with respect to the parameter space \cite{Gill00}.
   Eq.(\ref{eq:gCRineq}) indicates that for sufficiently large $N$, the risk
   function can decrease at most inverse-proportionally to $N$, and the rate is
   characterized by $\mbox{tr}[H_{\bm{s}}F_{\bm{s}}^{-1}]$.
   We can rewrite this as
   \begin{eqnarray}
      \mbox{tr}[H_{\bm{s}}F_{\bm{s}}^{-1}]
      &= \mbox{tr}[\sqrt{H_{\bm{s}}}F_{\bm{s}}^{-1}\sqrt{H_{\bm{s}}}]\\
      &= \sum_{\alpha=1}^{k}\sigma_{\alpha}(G_{\bm{s}}),
   \end{eqnarray}     
   where $\sigma_{\alpha}(A)$ is the $\alpha$-th eigenvalue of a symmetric $k\times
   k$ matrix $A$ arranged in decreasing order.
   Therefore, the rates of decrease of error probability
   and risk function are both characterized by, respectively, the
   maximal eigenvalue and the sum of all the eigenvalues of a common
   matrix $G_{\bm{s}}$.
   The rates' properties depend upon the choice of the loss function.
   For example, when we choose the Kullback-Leibler divergence,
   {\it i.e.}, $\Delta (\bm{s},\bm{s}^{\prime})=K(p_{\bm{s}}\|
   p_{\bm{s}^{\prime}})$, we obtain $H_{\bm{s}}=F_{\bm{s}}$ and
   therefore $\sigma_{1}(G_{\bm{s}})=1$ and
   $\sum_{a=1}^{k}\sigma_{a}(G_{\bm{s}})=k$.
   In this case the rates of decrease do not depend upon the true
   parameter or the tester, but in general the rates depend upon both.
   
   The Cram\'{e}r-Rao inequality holds only for unbiased estimators, and the
   bound can be broken by biased estimators.
   On the other hand, the error probability inequality holds for any consistent
   estimator.
   A maximum likelihood estimator is consistent under some conditions
   (including the identifiability condition),
   and is not unbiased in general but achieves the lower bound of Eq.(\ref{eq:gCRineq}) asymptotically.
   When we use a maximum likelihood reconstruction scheme in quantum
   tomography, the performance of the tester is evaluated by
   $\sum_{\alpha=1}^{k}\sigma_{\alpha}(G_{\bm{s}})$ from the risk function viewpoint.
   When we have two testers with the same value of
   $\sum_{\alpha=1}^{k}\sigma_{\alpha}(G_{\bm{s}})$ at a $\bm{s} \in S$,
   their performances are equivalent in the risk function sense,
   but if the maximal eigenvalues $\sigma_{1}(G_{\bm{s}})$ are different,
   their error probability performances are different.
   Thus we can evaluate the performance of testers more discerningly by
   considering error probabilities than we can by considering only risk functions,
   using the same set of eigenvalues -- that of the matrix $G_{\bm{s}}$.

   \subsection{Example}\label{subsec:example}
   
\begin{figure}[htbp]
   \includegraphics[width=60mm]{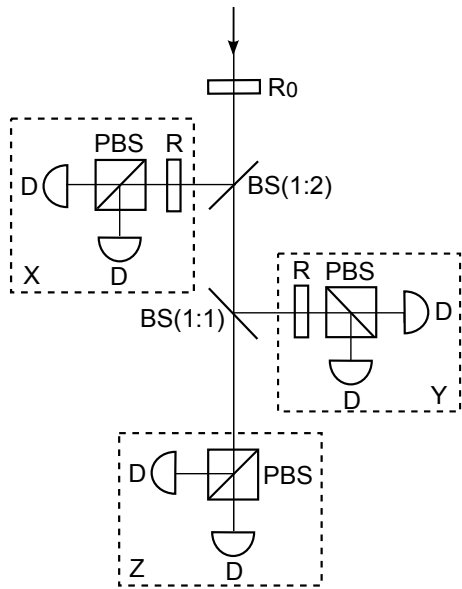}%
   \caption{\label{fig1}An experimental realization of a 6-state POVM in a photon polarization experiment, consisting of photodetectors (D), beam splitters (BS), polarizing beam splitters (PBS), and rotators (R). The rotator $\mathrm{R}_{0}$ defines the direction of the z-projective measurement, and the angles of rotators in X and Y are suitably chosen.}
   \end{figure}

   Here we analyze a simple example of a tester in 1-qubit state tomography; a 6-state POVM
\begin{eqnarray}   
   \bm{\Pi}=\{\frac{1}{3}|\uparrow_{x}\rangle \langle \uparrow_{x}|,\frac{1}{3}|\downarrow_{x}\rangle \langle \downarrow_{x}|,\frac{1}{3}|\uparrow_{y}\rangle \langle \uparrow_{y}|,\notag \\  \frac{1}{3}|\downarrow_{y}\rangle \langle \downarrow_{y}|,\frac{1}{3}|\uparrow_{z}\rangle \langle \uparrow_{z}|,\frac{1}{3}|\downarrow_{z}\rangle \langle \downarrow_{z}|\}.
\end{eqnarray}   
    This is constructed by mixing the $x$-, $y$-, and $z$-projective measurements randomly, as in Fig.\ref{fig1}. 
   This example will serve to illustrate how the performances of risk function and error probability approaches can differ; see the discussion in subsection \ref{subsec:evaluating_tester_performance}.
   
\begin{figure*}[htbp]
   \begin{minipage}{0.9\linewidth}
   \includegraphics[width=\linewidth]{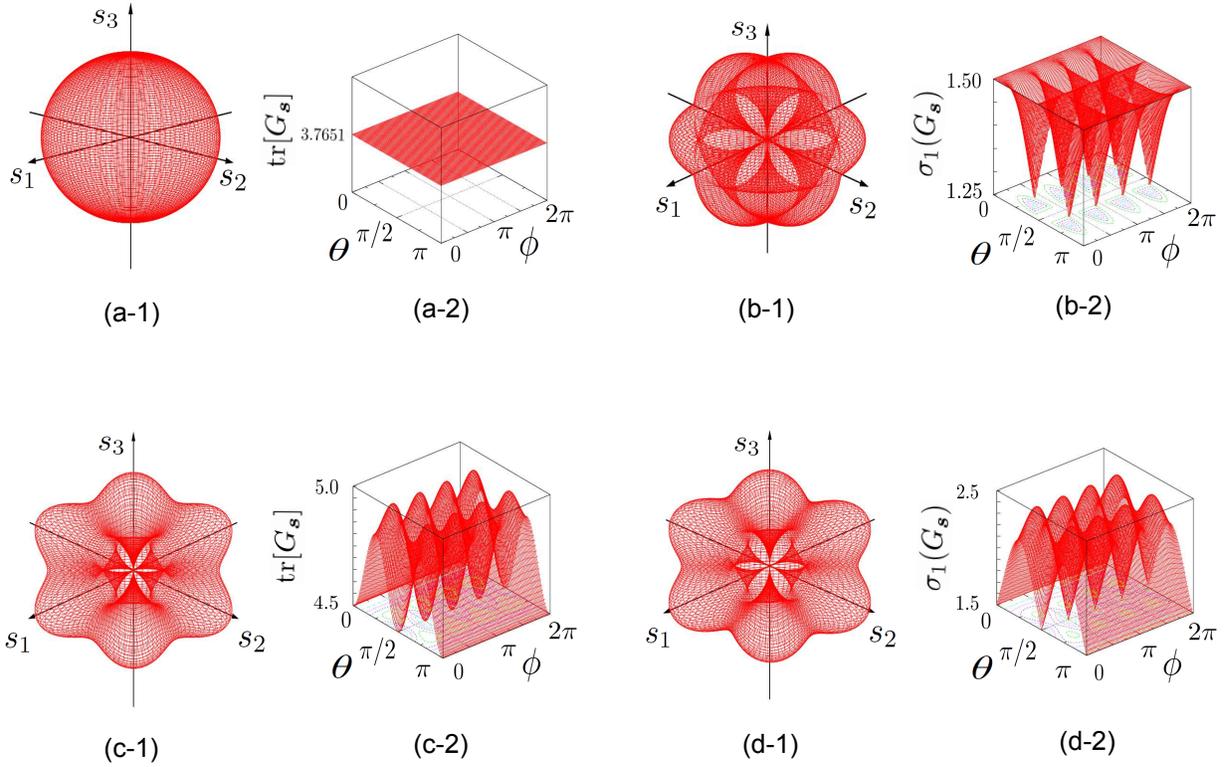}
   \end{minipage}
   \caption{\label{fig2}The dependency of the rates of decrease of risk function and error probability at $\| \bm{s}\|=0.7$ against $\theta$ and $\phi$:  $\mbox{tr}[G_{\bm{s}}]$ for the Hilbert-Schmidt distance (a-1) and (a-2) corresponding to Eq.(\ref{eq:GHS1}), and for the fidelity distance (c-1) and (c-2) corresponding to Eq.(\ref{eq:GF1}).  $\sigma_{1}(G_{\bm{s}})$ for the Hilbert-Schmidt distance (b-1) and (b-2) corresponding to Eq.(\ref{eq:GHS2}), and for the fidelity distance (d-1) and (d-2) corresponding to Eq.(\ref{eq:GF2}). These figures show that the bounds for risk function and error probability depend on the choice of the loss function.}
   \end{figure*}

   We choose a Bloch parametrization of the unknown state $\hat{\rho}(\bm{s})=\frac{1}{2}(\hat{I}+\bm{s}\cdot \hat{\bm{\sigma}})$.
   Then the inverse of the Fisher matrix is found to be
   \begin{eqnarray}
      F_{\bm{s}}^{-1}= 3\left(
      \begin{array}{ccc}
      1-(s_{1})^2 & 0 & 0\\
      0 & 1-(s_{2})^2 & 0\\
      0 & 0 & 1-(s_{3})^2
      \end{array}
      \right).\label{eq:F}
   \end{eqnarray}
   As the first example, we choose the square of the Hilbert-Schmidt distance $\Delta^{\mathrm{HS}}(\bm{s},\bm{s}^{\prime})^2:=\mbox{Tr}[(\hat{\rho}(\bm{s})-\hat{\rho}(\bm{s}^{\prime}))^{2}]$ and the square of the trace distance $\Delta^{\mathrm{T}}(\bm{s},\bm{s}^{\prime})^2:=\frac{1}{4}\mbox{Tr}[|\hat{\rho}(\bm{s})-\hat{\rho}(\bm{s}^{\prime})|]^2$ as the loss functions. Then we obtain that $\Delta^{\mathrm{HS}}(\bm{s},\bm{s}^{\prime})^2=\Delta^{\mathrm{T}}(\bm{s},\bm{s}^{\prime})^2=\frac{1}{4}\|\bm{s}-\bm{s}^{\prime}\|^2$. The Hesse matrix $H_{\bm{s}}^{\mathrm{HS}}(=H_{\bm{s}}^{\mathrm{T}})$ is $\frac{1}{2}I$, and the modified information matrix is $G_{\bm{s}}^{\mathrm{HS}}=G_{\bm{s}}^{\mathrm{T}}=\frac{1}{2}F_{\bm{s}}^{-1}$. We obtain
   \begin{eqnarray}
      \mbox{tr}[G_{\bm{s}}^{\mathrm{HS}}] &=& \mbox{tr}[G_{\bm{s}}^{\mathrm{T}}] = \frac{3}{2}(3-\|\bm{s}\|^2),\label{eq:GHS1}\\
      \sigma_{1}(G_{\bm{s}}^{\mathrm{HS}}) &=& \sigma_{1}(G_{\bm{s}}^{\mathrm{T}})\notag \\
         &=& \frac{3}{2}(1-\mbox{min}\{(s_{1})^2,(s_{2})^2,(s_{3})^2\}).\label{eq:GHS2}
   \end{eqnarray}
   We can readily see that
   \begin{eqnarray}
   3\le \mbox{tr}[G_{\bm{s}}^{\mathrm{HS}}] = \mbox{tr}[G_{\bm{s}}^{\mathrm{T}}] \le \frac{9}{2},\\
   \frac{3}{2}-\frac{\|\bm{s}\|^2}{2}\le \sigma_{1}(G_{\bm{s}}^{\mathrm{HS}}) = \sigma_{1}(G_{\bm{s}}^{\mathrm{T}}) \le \frac{3}{2},
   \end{eqnarray}
   where the lower bound of the maximal eigenvalue is achieved at the points satisfying $|s_1|=|s_2|=|s_3|=\frac{\|\bm{s}\|}{\sqrt{3}}$.
   Eq.(\ref{eq:GHS1}) indicates that the rate of decrease of the risk function depends only on the radius $r = \| \bm{s}\|$ of the Bloch vector and is independent of the angles $\theta$ and $\phi$. On the other hand, Eq.(\ref{eq:GHS2}) indicates that the rate of decrease of the error probability depends on all parameters $r,\ \theta,\ \phi$ (Fig.\ref{fig2} (a-1), (a-2), (b-1), (b-2)).  
   
   Next, we choose a squared fidelity distance $\Delta^{\mathrm{F}}(\bm{s},\bm{s}^{\prime})^2:=1-f(\bm{s},\bm{s}^{\prime})^2$ as the loss function, where $f(\bm{s},\bm{s}^{\prime})$ is the fidelity between $\hat{\rho}(\bm{s})$ and $\hat{\rho}(\bm{s}^{\prime})$. In 1-qubit case, the square of the fidelity is written as \cite{Bagan06-2}
   \begin{eqnarray}
     \! \! \! f(\bm{s},\bm{s}^{\prime})^2 = \frac{1}{2}(1+\bm{s}\cdot \bm{s}^{\prime}+\sqrt{(1-\|\bm{s}\|^2)(1-\| \bm{s}^{\prime} \|^2)}.\label{eq:fidelity}
   \end{eqnarray}
   We can calculate the Hesse matrix of $\Delta^{\mathrm{F}}$ and the root square from Eq.(\ref{eq:fidelity}) as 
   \begin{eqnarray}
      H_{\bm{s}}^{\mathrm{F}} &=& \frac{1}{2}(I+\frac{\bm{s}\bm{s}^{T}}{1-\|\bm{s}\|^2}),\label{eq:HF}\\
      \sqrt{H_{\bm{s}}^{\mathrm{F}}} &=& \frac{1}{\sqrt{2}}\{ I+ (\frac{1}{\sqrt{1-\|\bm{s}\|^2}}-1)\frac{\bm{s}\bm{s}^{T}}{\|\bm{s}\|^2}\}.
   \end{eqnarray}
   From Eq.(\ref{eq:F}) and Eq.(\ref{eq:HF}), we obtain 
   \begin{eqnarray}
   \! \! \! \! \!  \! \! \! \! \! \!  \mbox{tr}[G_{\bm{s}}^{\mathrm{F}}] = \frac{9}{2}+\frac{3}{1-\|\bm{s}\|^2}\{(s_{1}s_{2})^{2}+(s_{2}s_{3})^{2}+(s_{3}s_{1})^{2}\}.\label{eq:GF1}
   \end{eqnarray}
   and
   \begin{eqnarray}
      \sigma_{1}(G_{\bm{s}}^{\mathrm{F}}) = \sigma_{1}(\sqrt{H_{\bm{s}}^{\mathrm{F}}}F_{\bm{s}}^{-1}\sqrt{H_{\bm{s}}^{\mathrm{F}}})\label{eq:GF2}.
   \end{eqnarray}
   Eq.(\ref{eq:GF1}) indicates that the rate of decrease of the risk function for the fidelity distance depends on all parameters $r,\ \theta,\ \phi$, with plots given in Fig.\ref{fig2} (c-1), (c-2). The calculation of the largest eigenvalue $\sigma_1(G_{\bm{s}}^{\mathrm{F}})$ is done numerically, with results plotted in Fig.\ref{fig2} (d-1), (d-2). The figures in Fig.\ref{fig2} indicate that the rates of decrease of risk function and error probability change dramatically with the choice of the loss function.

   \subsection{Extension to more general quantum estimation problem}\label{subsection:general}

   A loss function used in quantum state tomography is usually a distance on $S$
   (or $\mathcal{S}(\mathcal{H})$).
   This is because the purpose of quantum state tomography is to identify
   the true parameter (or true density operator).
   There are, however, cases where exact identifiability is not
   required, for example, estimations of the average value of an Hermitian
   operator, the purity of an unknown state, or the value of an entanglement measure.
   These examples correspond to the case where $g$ is a map from $S$ to
   $\mathbf{R}$.
   More generally, we can consider $g: S \to \mathbf{R}^{l}, l\le k=d^2 -1$.
   Theorem \ref{theorem:main} can be generalized to this case by modifying
   the identifiability condition (see Appendix) and changing the meaning of the
   superscript $-1$ from the inverse matrix to the Moore-Penrose generalized inverse.
   Specifically, when $l=1$ and the loss function $\Delta$ is the squared absolute
   value, {\it i.e.}, $g: S \to \mathbf{R}$ and $\Delta(\bm{s},
   \bm{s}^{\prime})=|g(\bm{s})-g(\bm{s}^{\prime})|^2$, we can obtain
   \begin{eqnarray}
      H_{\bm{s}} = 2(\nabla_{\bm{s}}g)(\nabla_{\bm{s}}g)^{T},\\
      \frac{1}{\sigma_{1}(G_{\bm{s}})} = \frac{1}{2 \nabla_{\bm{s}}g \cdot
    F_{\bm{s}}^{-1} \nabla_{\bm{s}}g}.
   \end{eqnarray}   
   This result exactly coincides with the known result of Eq.(\ref{eq:r1}).   
     
   When the parameter space is 1-dimensional,
   the rates of decrease of the two evaluation methods are characterized
   by the same function, 
   but when the parameter space is more than 2-dimensional, the rates
   can be characterized differently.
   The most simple tomographic object, a 1-qubit
   state, has a 3-dimensional parameter space, therefore even in the simplest type of
   quantum tomography, if two given testers have the same rate of decrease of a risk
   function, their rates of decrease of error probability can be different,
   {\it i.e.}, the testers can have different quantum tomographic performance (see subsection \ref{subsec:evaluating_tester_performance}).

\section{Discussion}\label{section:discussion}

   \subsection{Evaluating tester performance}\label{subsec:evaluating_tester_performance}

   Our result shows that when the true parameter is $\bm{s}$,
   the rate of decrease of the error probability is characterized by
   $\sigma_{1}(G_{\bm{s}})$. 
   In real experiments of course, we do not know the true parameter, which is the
   reason we perform tomography in the first place.
   We explain three approaches to evaluating tester performance below.
   
   The first is to use a parameter which we expect as the true parameter.
   In many experiments, quantum state tomography is performed not for
   estimating a state but for proving an
   experimental realization of a specific quantum state, for example, a
   maximally entangled state.
   By using the parameter corresponding to the quantum state we want
   to realize, we can evaluate the tester's performance in achieving that state.
   Of course the disadvantage of this method is that this evaluation result can be
   different from the true performance in the experiment, because the true
   parameter can be different from the parameter which we expect.
   
   The second is to consider the average performance.
   Let $\mu$ denote a measure on the parameter space $S$.
   We define the average performance of the error probability with
   respect to a measure $\mu$ as
   \begin{eqnarray}
      \int_{S}d\mu(\bm{s})\sigma_{1}(G_{\bm{s}}).
   \end{eqnarray}      
   In this approach, a tester with smaller average rate of decrease is
   better.
   The average performance can be calculated without knowing the true
   parameter, but of course it is not guaranteed that the average value is
   equivalent to the true performance in the experiment.
   Since this evaluation results depend upon the choice of the measure $\mu$,
   we need to ascertain the validity of the choice.
   
   The third is to consider the worst case performance.
   We define the worst case performance of a tester as
   \begin{eqnarray}
      \max_{\bm{s}\in S}\sigma_{1}(G_{\bm{s}}).
   \end{eqnarray}    
   This can be calculated without the true parameter,
    and it is guaranteed that the true performance is
   necessarily better or equal to the value.
   The disadvantage of this method is that we might evaluate the tester's
    performance much lower than the true performance in the experiment.

   As an example we compare the performance of testers according to the first approach.
   Let us consider a 6-state POVM explained in \ref{subsec:example} and Fig.\ref{fig1}.
   Suppose that the density operator which we try to realize is characterized by $(r=0.7,\ \theta=0,\ \phi=0)$.
   The rates of decrease of risk function and error probability for $\Delta^{\mathrm{HS}},\Delta^{\mathrm{T}},\Delta^{\mathrm{F}}$ are characterized by $\mbox{tr}[G_{\bm{s}}]$ and $\sigma_{1}(G_{\bm{s}})$, given in Eqs.(\ref{eq:GHS1}), (\ref{eq:GHS2}), (\ref{eq:GF1}), (\ref{eq:GF2}), and Fig.\ref{fig2}.
   Suppose we tune the angles $\theta_{0}$ and $\phi_{0}$ of the rotator $\mathrm{R}_{0}$ in Fig.\ref{fig1}. Then the true Bloch vector is rotated to $(r=0.7,\ \theta=\theta_{0},\ \phi=\phi_{0})$.
   Which angles $\theta_{0}$ and $\phi_{0}$ should we choose for the state tomography?
   The true density operator may not be what we want, but it is expected to that we want because we make effort to realize the state in the experiment.
   So, it is natural to tune the angle $\theta_{0}$ and $\phi_{0}$ so that the statistical error becomes as small as possible at the rotated objective density operator $(r=0.7,\ \theta=\theta_{0},\ \phi=\phi_{0})$.   
   
   If we use the square of the Hilbert-Schmidt distance as the loss function, the rate of decrease of the risk function is independent of the angle of the rotator (Fig.\ref{fig2} (a-1), (a-2)). Experimental setups with any angle of $\mathrm{R}_{0}$ have equivalent performance from the risk function viewpoint.
   On the other hand, the rate of decrease of the error probability depends on those angles (Fig.\ref{fig2} (b-1), (b-2)). We should tune the angle to the point where $(r=0.7,\ \theta=\theta_{0},\ \phi=\phi_{0})$ is at one of the minima in Fig.\ref{fig2} (b-2). 
   Our error probability approach therefore allows us to evaluate the statistical performance of these testers (experiments with varying the angles of $\mathrm{R}_0$) while a risk function approach would not. 
   If we use the fidelity distance, the minima of the risk function and the error probability are the same, (although the curves are not, as the figures show), and we should choose the angle such that $(r=0.7,\ \theta=\theta_{0},\ \phi=\phi_{0})$ is at one of  the minima in Fig.\ref{fig2} (c-2) and (d-2).  This illustrates that the difference between the approaches hinges upon the choice of loss function we use in our analysis.

   \subsection{Extension to infinite sample space}\label{subsection:infinite_sample_space}

      Theorem \ref{theorem:main} holds for a finite sample space.
   For a specific case ($g:S\to \mathbf{R}$ and $\Delta$ is the squared absolute
   value),
   it is known that Eq.(\ref{eq:BGZ}) also holds for infinite sample space under some
   regularity conditions \cite{Bahadur1960, Bahadur1967}.
   We can prove that Theorem \ref{theorem:main} holds for infinite sample
   space under some conditions by combining the proof in
   \cite{Bahadur1960, Bahadur1967} with Sanov's theorem and using
   the linear parametrizability of probability distributions in quantum
   mechanics.
   Therefore, Theorem \ref{theorem:main} holds not only for finite, but
   also infinite sample spaces.
   However, any real experiments will have finite detector resolution,
   and so finite sample spaces suffice.
   
   \subsection{Effect of parameter space boundary}

   In Theorem \ref{theorem:main}, the true parameter is limited to the interior $S^{o}$.
   Hence it cannot be applied to parameters on the
   boundary $\partial S:= S\setminus S^{o}$ which corresponds to the set of
   all non full rank density operators, including all pure states.
   This limitation can be overlooked by invoking decoherence: in real experiments
   the system of interest is uncontrollably affected by the enviroment,
   leading to full rank states parametrized in the interior.
   The reason behind the limitation is very technical,
   stemming from the fact that in our proof we assume the invariance of the support of probability distribution,
   differentibility, and openness at each point of the parameter space.
   Such regularity conditions are assumed in
   standard classical statistical estimation theory.
   Statistical models that do not satisfy the regularity conditions are called non-regular,
   and it is known that they can behave very
   differently from regular statistical models \cite{Akahara-Takeuchi95}.    
   The analysis of risk functions and error probabilities at $\partial S$
   is an open problem.

   \subsection{Relation to quantum Fisher matrix}\label{subsection:quantum_estimation_theory}

   There is an approach to statistical estimation in quantum systems using a quantity called the quantum Fisher matrix. In this subsection, we briefly explain the relationship between quantum and classical Fisher matrix approaches.
   
   The quantum Fisher matrix approach is an attempt to derive the maximal value of the information extractable from a quantum system. The quantum Fisher matrix is defined as the matrix satisfying 
   \begin{eqnarray}
      F_{\bm{s}}^{Q} \ge F_{\bm{s}}(\bm{\Pi}),\label{eq:QFisher}
   \end{eqnarray}
   for all POVMs $\bm{\Pi}$ and $\bm{s}\in S$, where $F_{\bm{s}}(\bm{\Pi})$ is the usual Fisher matrix,
   as well as a monotonicity condition under quantum operations \cite{Helstrom76,Holevo82,Hayashi05}.  We put $(\bm{\Pi})$ in order to emphasize the dependency on the POVM. 
   By combining Eq.(\ref{eq:gCRineq}) and Eq.(\ref{eq:QFisher}), we can obtain the quantum Cram\'{e}r-Rao inequality,
   \begin{eqnarray}
      \bar{\Delta}^{(N)} \ge \frac{\mbox{tr}[H_{\bm{s}}F_{\bm{s}}^{Q-1}]}{2N} + o(\frac{1}{N}). \label{eq:QCR} 
   \end{eqnarray}
   By definition, the quantum Fisher matrix depends only on the true density operator and is independent of POVMs.  So from the risk function viewpoint, the quantum Fisher matrix can be interpreted as the principal bound of the rate of decrease for a fixed true density operator. 
   By combining our result, Eq.(\ref{eq:main1}) and Eq.(\ref{eq:QFisher}), we can obtain 
   \begin{eqnarray}
      \varliminf_{\epsilon \to 0}\varliminf_{N\to \infty}\frac{1}{\epsilon^2 N}\log
   P^{(N)}_{\bm{s}}(\Delta(\bm{s}^{\mathrm{est}}_{N}, \bm{s})>
   \epsilon^2 ) \notag \\
   \ge -1/\sigma_{1}( \sqrt{H_{\bm{s}}}F_{\bm{s}}^{Q-1}\sqrt{H_{\bm{s}}}).\label{eq:QBahadur}
   \end{eqnarray}
   In general, however, there are no POVMs achieving the equality in Eq.(\ref{eq:QFisher}), except for specific cases which include one dimensional parameter space \cite{Hayashi05}.  So the bound is not tight in general multi-parameter estimation, like quantum tomography.
   We use the classical Fisher matrix here because we are interested in evaluating the performance of a fixed experimental apparatus (tester), and we therefore require POVM dependence.  One could evaluate the performance of a POVM by comparing the value of $\sigma_{1}( \sqrt{H_{\bm{s}}}F_{\bm{s}}^{-1}\sqrt{H_{\bm{s}}})$ with $\sigma_{1}( \sqrt{H_{\bm{s}}}F_{\bm{s}}^{Q-1}\sqrt{H_{\bm{s}}})$, but the compared bound is not achievable in general. The derivation of the optimal POVM is an open problem.

\section{Summary}\label{section:summary}

   In this paper, we proved a large deviation inequality for consistent
   estimators in quantum tomography by using classical statistical
   estimation techniques. 
   The inequality shows that, under some conditions, the error probability of any
   consistent estimator can decrease at most exponentially with respect to the total number
   of measurement trials, and there is a bound of the rate of decrease
   which is achievable by a maximum likelihood estimator under the
   informational completeness of the tester.
   We also derived the explicit form of the bound and
   proved that known quantum tomography conditions are equivalent to the
   identifiability condition in classical estimation theory.  
   
   From our results, it is shown that a risk function and error probability
   measured by the same loss function are characterized by a common matrix,
   the inverse Fisher matrix modified by the loss function.
   The rate of decrease (with respect to the number of trials) of the
   risk function is characterized by the sum of the
   eigenvalues of this matrix, and that of the error probability 
   by the maximal eigenvalue.
   The Cram\'{e}r-Rao inequality, which is a known risk function inequality,
   holds only for unbiased estimators, and the
   bound can be broken by biased estimators.
   On the other hand, the error probability inequality holds for any consistent
   estimator which gives us the true object in the limit of infinite trials.
   Therefore, the lower bound of the error probability characterizes the performance of the given apparatus, independently of the choice of estimator.  
   The explicit form of the bound makes it possible to quantify the performance of the
   apparatus for the estimation purpose in the error probability sense.
   We showed, by using a 6-state POVM in single qubit state tomography as an example, that
   by combining our error probability approach with a risk function approach,
   we can evaluate the performance more discerningly than we can by considering only risk functions.

\appendix*
%\section{}
\section{Proof of main theorem}\label{section:appendix}

We give the detailed proof of Theorem 1, using classical
statistical estimation theory.
We divide the proof into six parts in order to clarify the role of each
condition, as well as to isolate the role of
quantum mechanics in the main result.

   \subsection{Six lemmas}

   We first consider the setup described in section
   \ref{section:classical_estimation_theory}, that is, we do not assume
   the statistical model given by Eq.(\ref{eq:p=rhopi}).
   Suppose that the parameter space $\Theta$ is a closed compact subset
   of $\mathbf{R}^k$.
   Let $\partial \Theta$ denote the boundary of $\Theta$, that is, $\partial
   \Theta:=\Theta\setminus \Theta^{o}$ and assume that $\Theta^{o}$ is
   open and nonempty.
   We also assume that $p_{\theta}(x)$ is a thrice differentiable function with respect to
          $\theta\in \Theta$ for any $x \in \Omega$.
   Note that these assumptions are satisfied in quantum mechanics for
   finite dimensional systems.

% Lemma 1
   First, we prove that Eq.(\ref{eq:BGZ}) holds for any estimator
   consistent not only in distances, but also in pseudo-distances.
   \begin{lemma}\label{lemma:1}
      Suppose that $\Delta$ is a pseudo-distance on $\Theta$.
      If $\theta \in \Theta^{o}$, for an arbitrary consistent estimator
    $\theta^{\mathrm{est}}$ in $\Delta$,
      the following inequality holds:
      \begin{eqnarray}
         \varliminf_{N \to \infty}\frac{1}{N} \log 
         P_{\theta}^{(N)}(\Delta (\theta^{\mathrm{est}}_{N}, \theta)>\epsilon^2 )\notag \\
         \ge -\inf_{\theta^{\prime} \in \Theta}\{ K(p_{\theta^{\prime}}\| p_{\theta}); \Delta (\theta^{\prime}, \theta)>\epsilon^2 \}
      \end{eqnarray}
   \end{lemma}
   {\bf Proof:}
   This is a straightforward generalizations of the proof in \cite{BGZ1980}, so we omit it here. 
   $\square$

   From Lemma 1,  we obtain  
   \begin{eqnarray}
      \varliminf_{\epsilon\to 0}\varliminf_{N\to \infty}\frac{1}{\epsilon^2 N}\log
    P^{(N)}_{\theta}(\Delta(\theta^{\mathrm{est}}_{N}, \theta)>\epsilon^2 ) \notag \\
      \ge -\varlimsup_{\epsilon\to 0}\frac{1}{\epsilon^2}\inf_{\theta^{\prime} \in
     \Theta}\{ K({p}_{\theta^{\prime}} \| {p}_{\theta}); 
    \Delta(\theta^{\prime}, \theta)>\epsilon^2 \}.\label{eq:A2}
   \end{eqnarray}

% Lemme 2
   Second, we introduce a lemma for calculating the R.H.S. of Eq.(\ref{eq:A2}).
   \begin{lemma}\label{lemma:2}
   Let $A$ and $B$ be $k\times k$ real, positive-semidefinite matrices.
   If $\mathrm{supp}A \supseteq \mathrm{supp}B$ holds, then
   \begin{eqnarray}
      \inf_{\bm{a}\notin \mathrm{ker}B}\{ \frac{\bm{a}\cdot A \bm{a}}{\bm{a}\cdot B \bm{a}}\} 
       = \frac{1}{\sigma_{1}(\sqrt{B}A^{-1}\sqrt{B})}\label{eq:inf-AB}
   \end{eqnarray}
   holds where $A^{-1}$ is the Moore-Penrose generalized inverse of $A$. \label{lemma:abc-c}
   \end{lemma}
     {\bf Proof:}  
     Let us define $\bm{b}:=\sqrt{A}\bm{a}/\|\sqrt{A}\bm{a} \|$.
     Then,
     \begin{eqnarray}
        &&\inf_{\bm{a}\notin \mathrm{ker}B}\{ \frac{\bm{a}\cdot A \bm{a}}{\bm{a}\cdot B \bm{a}}\}\notag \\ 
        &=& \inf_{\bm{b}\notin \mathrm{ker}\sqrt{A}^{-1}B\sqrt{A}^{-1};\|\bm{b}\|=1}\frac{1}{ \bm{b}\cdot
        \sqrt{A}^{-1}B\sqrt{A}^{-1}\bm{b} } \\
        &=& 1/\sigma_{1}(\sqrt{A}^{-1}B\sqrt{A}^{-1}).
     \end{eqnarray}
     Let us consider the singular value decomposition of $\sqrt{A}^{-1}\sqrt{B}$, {\it i.e.}, $\sqrt{A}^{-1}\sqrt{B}=U_{1}\Lambda U_{2}$, where $U_{1}$ and $U_{2}$ are $k\times k$ unitary matrices and $\Lambda$ is a diagonalized matrix.
     We obtain
     \begin{eqnarray}
        \sqrt{A}^{-1}B\sqrt{A}^{-1} 
        &=& (\sqrt{A}^{-1}\sqrt{B}) (\sqrt{A}^{-1}\sqrt{B})^{T}\\
        &=& U_{1}\Lambda^2 U_{1}^{T},\\
        \sqrt{B}A^{-1}\sqrt{B}
        &=& (\sqrt{A}^{-1}\sqrt{B})^{T} (\sqrt{A}^{-1}\sqrt{B})\\
        &=& U_{2}^{T}\Lambda^2 U_{2}.
     \end{eqnarray}
     Therefore $\sigma_{1}(\sqrt{A}^{-1}B\sqrt{A}^{-1})=\sigma_{1}(\sqrt{B}A^{-1}\sqrt{B})$.
     $\square$\\
     Note that when $A$ is full rank, the Moore-Penrose generalized inverse coincides with the (usual) inverse.

% Lemma 3
   Third, we calculate the infimum on the R.H.S. of Eq.(\ref{eq:A2}).
   \begin{lemma}\label{lemma:3}
   Suppose that $\Delta$ is a sufficient smooth pseudo-distance with a
    non-zero same point Hesse matrix $H_{\theta}$.
   Then
   \begin{eqnarray}
   \varlimsup_{\epsilon\to 0}\frac{1}{\epsilon^2}\inf_{\theta^{\prime} \in \Theta}\{
    K({p}_{\theta^{\prime}} \| {p}_{\theta}); 
    \Delta(\theta^{\prime}, \theta)>\epsilon^2 \}\notag \\
    = \frac{1}{\sigma_{1}\bigl( \sqrt{H_{\theta}}F^{-1}_{\theta}\sqrt{H_{\theta}} \bigr)}.\label{eq:A.8}
   \end{eqnarray}
   holds.
   \end{lemma}
   {\bf Proof:}
   Let us define $B(\theta^{\prime}, \theta):=2 \frac{\Delta(\theta^{\prime}, \theta)}{\|
 \theta^{\prime} - \theta\|^2}$.
   Then 
   \begin{eqnarray}
      \hspace{-7mm} B(\theta^{\prime},\theta) = \frac{(\theta^{\prime}
    -\theta)}{\|\theta^{\prime}-\theta\|}\cdot H_{\theta}
    \frac{(\theta^{\prime} -\theta)}{\|\theta^{\prime}-\theta\|}+
    O(\| \theta^{\prime} - \theta\|),
   \end{eqnarray}
   and the first term is independent of $\|\theta^{\prime}-\theta\|$.
   Then, for sufficiently small $\epsilon$,
   \begin{eqnarray}
    &&\frac{1}{\epsilon^2}\inf_{\theta^{\prime}\in \Theta}\{ K({p}_{\theta^{\prime}}\| {p}_{\theta});
    \Delta (\theta^{\prime}, \theta)>\epsilon^2 \} \notag \\
      &=& \frac{1}{\epsilon^2}\inf_{\theta^{\prime}\in \Theta} \{ K(p_{\theta^{\prime}}\| p_{\theta}) ; \| \theta^{\prime} -\theta\| >\! \! \epsilon\sqrt{ \frac{2}{B(\theta^{\prime},\theta)} } \} \\ 
      &=& \frac{1}{\epsilon^2}\inf_{\theta^{\prime}\in \Theta} \{
    \frac{1}{2}(\theta^{\prime}-\theta)F_{\theta}(\theta^{\prime}-\theta)+O(\|\theta^{\prime}-\theta 
    \|^3) ; \notag \\
      & & \|\theta^{\prime} -\theta \|>\epsilon\sqrt{ \frac{2}{B(\theta^{\prime},\theta)}} \}\\
%      &=& \frac{1}{\epsilon}\inf \{ \frac{2\epsilon}{B(\theta^{\prime},\theta)}
%    \frac{1}{2}{\nu}\cdot F_{\theta}{\nu} |\ {\nu} : \|{\nu}\|=1, {\nu}\cdot H_{\theta}{\nu}>0 \}\\
    &=& \inf_{ \bm{a}\notin \mathrm{ker}H_{\theta}} \{ \frac{\bm{a}\cdot F_{\theta}{\bm{a}}}{\bm{a}\cdot
     H_{\theta}\bm{a}}; \|\bm{a}\|=1 \}\\
    &=& \frac{1}{\sigma_{1}\bigl( \sqrt{H_{\theta}}F^{-1}_{\theta}\sqrt{H_{\theta}} \bigr)},
   \end{eqnarray}
   where we used Lemma 2 in the last line.
   Note that Eq.(\ref{eq:A.8}) holds not only for the linit superior
   $\varlimsup_{\epsilon \to 0}$, but
   also for the limit inferior $\varliminf_{\epsilon \to 0}$.
   $\square$    

   From Lemma 1 and Lemma 3, we obtain the following inequality for any
   estimator consistent in a sufficiently smooth pseudo-distance with the Hesse matrix $H_{\theta}$: 
   \begin{eqnarray}
      \varliminf_{\epsilon\to 0}\varliminf_{N\to \infty}\frac{1}{\epsilon^2 N}\log
    P^{(N)}_{\theta}(\Delta(\theta^{\mathrm{est}}_{N}, \theta)>\epsilon^2 )\notag \\
      \ge - \frac{1}{\sigma_{1}\bigl( \sqrt{H_{\theta}}F^{-1}_{\theta}\sqrt{H_{\theta}} \bigr)}.\label{eq:Amain1}
   \end{eqnarray}

% Lemma 4
    Fourth, we prove that if the identifiability condition is satisfied,
   then a maximum likelihood estimator is consistent in the
   pseudo-distance $\Delta$.
   In preparation, we introduce empirical measures.
% An empirical measure 
   Given a finite sequence $\bm{x}^N = \{x_{1}, \ldots , x_{N} \}$ and
   $Y\in \mathscr{B}$,
   the empirical measure $L_{N}^{\bm{x}^N}$ induced by the sequence is defined as 
   \begin{eqnarray}
      L_{N}^{\bm{x}^N}(Y):=
    \sum_{y\in Y}\frac{1}{N}\sum_{i=1}^{N}\delta_{y,x_{i}},     
   \end{eqnarray}
   where $\delta_{y,x}$ is Kronecker's delta.
   Then the value of the empirical measure on an elemental set $\{x\}\in
   \mathscr{B}$ is equivalent to the relative frequency of $x$ for the
   data $\bm{x}^N$, {\it i.e.}, $f_{N}(x)= L_{N}^{\bm{x}^N}(\{x\})$.
   We identify $L^{\bm{x}^N}_{N}$ and $f_{N}$ below.

% Sanov's theorem
   Now we introduce Sanov's theorem for empirical measures.
   Let $P_{p}$ denote a probability measure on $\mathscr{B}$ with a
   probability distribution $p$.
   When $p\in \mathcal{P}_{\Theta}$, we have $P_{p}=P_{\theta}$.
   We use a notation $P_{p}^{(N)}(L_{N}^{X^N}\in 
   \mathcal{A}):=P_{p}^{(N)}(\{\bm{x}^N \in \Omega^N ; L_{N}^{\bm{x}^N}
   \in \mathcal{A}\})$, where $\mathcal{A}$ is a given set of probability
   distributions.
   \renewcommand{\thetheorem}{}
   \begin{theorem}[Sanov]
   For every set $\mathcal{A}$ of probability distributions in $\mathcal{P}(\Omega)$, 
   \begin{eqnarray}
    \hspace{-7mm}  -\inf_{p^{\prime} \in \mathcal{A}^{o}}K(p^{\prime}\| p) 
        &\le& \varliminf_{N\to\infty}\frac{1}{N}\log P_{p}^{(N)}(L_{N}^{X^N}\in
    \mathcal{A} )\\
        &\le& \varlimsup_{N\to\infty}\frac{1}{N}\log P_{p}^{(N)}(L_{N}^{X^N}\in
    \mathcal{A} )\\
        &\le& -\inf_{p^{\prime} \in \mathcal{A}}K(p^{\prime}\| p), 
   \end{eqnarray}
   where $\mathcal{A}^{o}$ is the interior of $\mathcal{A}$ considered as a subset
   of $\mathcal{P}(\Omega)$ and $K$ is the Kullback-Leibler divergence
   \cite{Sanov57, DemboZeitouni98}.
   \end{theorem}

   We are now in a position to prove the following lemma. 
   \begin{lemma}\label{lemma:4}
      If the identifiability condition is satisfied, then
    \begin{eqnarray} 
      \lim_{N\to \infty} P^{(N)}_{\theta}(\Delta
       (\theta^{\mathrm{ml}}_{N} , \theta) > \epsilon^2) = 0
    \end{eqnarray}
    holds for any $\epsilon >0$.  That is, a maximum
    likelihood estimator is consistent in a pseudo-distance $\Delta$ on $\Theta$.
   \end{lemma}
   {\bf Proof:}
% A maximum likelihood estimator
   A maximum likelihood estimate $\theta^{\mathrm{ml}}_{N}$ can be
   redefined by using the Kullback-Leibler divergence and the relative
   frequency as follows:
   \begin{eqnarray}
      \theta^{\mathrm{ml}}_{N} :&= \mbox{argmax}_{\theta\in\Theta}  \prod_{i=1}^{N}p_{\theta}(x_{i} )\\
      &= \mbox{argmin}_{\theta\in\Theta} K({f}_{N}\| {p}_{\theta}).
   \end{eqnarray}
   Let us define
    \begin{eqnarray}
      \theta_{{p}}:= \mbox{argmin}_{\theta\in\Theta} K({p}\| {p}_{\theta} ).  
    \end{eqnarray}
   Then $\theta^{\mathrm{ml}}_{N}= \theta_{{f}_{N}}$.
   When analyzing a maximum likelihood estimate $\theta^{\mathrm{ml}}_{N}$, we need to be careful to
   check whether $\theta^{\mathrm{ml}}_{N}$ is included in $\Theta^{o}$ or $\partial \Theta$.
   Let us introduce four sets of probability distributions $\mathcal{A}_{1}$, $\mathcal{A}_{2}$,
   $\mathcal{A}_{3}$, and $\mathcal{D}_{\theta, \epsilon}$ as
   \begin{eqnarray}
      \mathcal{A}_{1}:&=& \{{p} \in \mathcal{P}_{\Theta}; \theta_{p} \in \Theta^{o} \},\\
      \mathcal{A}_{2}:&=&\{{p} \in \mathcal{P}_{\Theta}; \theta_{p} \in \partial \Theta \},\\
      \mathcal{A}_{3}:&=& \mathcal{P}(\Omega)\setminus \mathcal{P}_{\Theta}, \\ 
      \mathcal{D}_{\theta, \epsilon}:&=& \{{p} \in \mathcal{P}(\Omega); \; \Delta(\theta_{p}, \theta)> \epsilon^2  \}.
   \end{eqnarray}
   If ${f}_{N} \in \mathcal{A}_{1} \cup \mathcal{A}_{2}(=\mathcal{P}_{\Theta})$, then ${p}_{\theta^{\mathrm{ml}}_{N}}={f}_{N}$.
   If ${f}_{N} \in \mathcal{A}_{3}$, then
   ${p}_{\theta^{\mathrm{ml}}_{N}}\in \mathcal{A}_{2}$ and
   ${p}_{\theta^{\mathrm{ml}}_{N}}\neq f_{N}$.
   Since $\mathcal{P}(\Omega)=\mathcal{A}_{1} \cup \mathcal{A}_{2} \cup
   \mathcal{A}_{3}$ and these sets are disjoint, we can rewrite the
   error probability as  
    \begin{eqnarray}
        &&P^{(N)}_{\theta}( \Delta (\theta^{\mathrm{ml}}_{N}, \theta) >
	 \epsilon^2)\notag \\
        &=& P^{(N)}_{\theta}({f}_{N} \in \mathcal{D}_{\theta, \epsilon})\\
        &=& P^{(N)}_{\theta}({f}_{N} \in \mathcal{A}_{1} \cap
	 \mathcal{D}_{\theta, \epsilon})
           + P^{(N)}_{\theta}({f}_{N} \in \mathcal{A}_{2} \cap
	 \mathcal{D}_{\theta, \epsilon})\notag \\
        & & + P^{(N)}_{\theta}({f}_{N} \in \mathcal{A}_{3} \cap \mathcal{D}_{\theta, \epsilon}).
    \end{eqnarray}
    
    Because $\Theta$ is compact and $\Theta^{o}$ is not empty, from Sanov's
     theorem, we can obtain 
    \begin{eqnarray}
       \lim_{n\to \infty}\frac{1}{N}\log P^{(N)}_{\theta}({f}_{N} \in
	\mathcal{A}_{j} \cap \mathcal{D}_{\theta, \epsilon})\notag \\
       = -\inf_{{p} \in \mathcal{A}_{j} \cap \mathcal{D}_{\theta, \epsilon}}
	K({p}\| {p}_{\theta}),\ j=1,2,3.
    \end{eqnarray}
    From the identifiability condition, 
    \begin{eqnarray}
      \inf_{p \in \mathcal{A}_{j}\cap \mathcal{D}_{\theta, \epsilon}} K(p \| p_{\theta}) > 0,\ j=1,2,3.
    \end{eqnarray}
    Therefore, for sufficiently large $N$, there exists $\nu$, $0<\nu <1$, such that
    \begin{eqnarray}
      P^{(N)}_{\theta}(\Delta (\theta^{\mathrm{ml}}_{N} , \theta) > \epsilon^2) < \nu^N
    \end{eqnarray}
    holds for any $\epsilon >0$.
    So, a maximum likelihood estimator is consistent in $\Delta$ under the
    identifiability condition. 
    $\square$

% Lemma 5   
    Fifth, we prove that if the identifiability condition is
    satisfied, a maximum likelihood estimator achieves the
    equality in Eq.(\ref{eq:Amain1}).
   \begin{lemma}\label{lemma:5}
      Suppose that $\Delta$ is a sufficiently smooth pseudo-distance on
    $\Theta$ with a non-zero same point Hesse matrix $H_{\theta}$.
    If the identifiability condition is satisfied, then
    \begin{eqnarray}
    \lim_{\epsilon \to 0}\lim_{N\to \infty}\frac{1}{\epsilon^2 N}\log
     P^{(N)}_{\theta}(\Delta (\theta^{\mathrm{ml}}_{N}, \theta)>\epsilon^2 )\notag \\
     = - \frac{1}{\sigma_{1}(\sqrt{H_{\theta}}F^{-1}_{\theta}\sqrt{H_{\theta}})}\label{eq:lemma5}
    \end{eqnarray}
    holds.
   \end{lemma}
   {\bf Proof:}
    From the continuity of $K$ and the openness of $S^{o}$,
    for arbitrary $\theta \in \Theta^{o}$, 
    there exists $\epsilon_{0} >0$ such that
    \begin{eqnarray}
       \inf_{{p} \in \mathcal{A}_{1} \cap \mathcal{D}_{\theta, \epsilon}} K({p} \|{p}_{\theta})
       < \inf_{{p} \in \mathcal{A}_{j}\cap \mathcal{D}_{\theta,
       \epsilon}} K({p} \| {p}_{\theta}),
    \end{eqnarray} 
    hold for $j=2,3$ and for any $\epsilon$ satisfying $0<\epsilon <\epsilon_{0}$ %\cite{Note3}.
    \footnote{The upper bound of $\epsilon_{0}$ depends on the true
    parameter $\theta$, and if $\epsilon$ is in $(0,\epsilon_{0})$, a
    maximum likelihood estimator achieves the equality in
    Eq.(\ref{eq:BGZ}) as pointed out in \cite{Kester86}. This is because the sample
    space is finite. If the sample space is infinite, the equality for
    arbitrary finite $\epsilon$ is not
    achievable by any estimate \cite{Rukhin83}.}.
    Hence, for sufficiently large $N$ and sufficiently small $\epsilon$, 
    \begin{eqnarray}
    \hspace{-7mm}P^{(N)}_{\theta}({f}_{N} \in \mathcal{A}_{1} \cap \mathcal{D}_{\theta, \epsilon})
           > P^{(N)}_{\theta}({f}_{N} \in \mathcal{A}_{j} \cap
	   \mathcal{D}_{\theta, \epsilon}),
    \end{eqnarray}
    hold for $j=2,3$, and we have
    \begin{eqnarray}
     && \varlimsup_{\epsilon \to 0}\lim_{N\to \infty}\frac{1}{\epsilon^2 N}\log
     P^{(N)}_{\theta}(\Delta (\theta^{\mathrm{ml}}_{N}, \theta)>\epsilon^2 ) \notag \\
     &=& \varlimsup_{\epsilon \to 0}\lim_{N\to \infty}\notag \frac{1}{\epsilon^2 N}\log
       \bigl[ P^{(N)}_{\theta}({f}_{N} \in \mathcal{A}_{1} \cap \mathcal{D}_{\theta, \epsilon})\notag \\ 
     & & \hspace{10mm}+ P^{(N)}_{\theta}({f}_{N} \in \mathcal{A}_{2}
      \cap \mathcal{D}_{\theta, \epsilon}) \notag \\
     & & \hspace{15mm}+ P^{(N)}_{\theta}({f}_{N} \in \mathcal{A}_{3} \cap \mathcal{D}_{\theta, \epsilon}) \bigr]\\
     &=& \varlimsup_{\epsilon \to 0}\lim_{N\to \infty}\frac{1}{\epsilon^2 N}\log
         P^{(N)}_{\theta}({f}_{N} \in \mathcal{A}_{1} \cap \mathcal{D}_{\theta, \epsilon})\\
     &=& \varlimsup_{\epsilon \to 0}\frac{1}{\epsilon^2}
         \bigl[- \inf_{{p} \in \mathcal{A}_{1}\cap \mathcal{D}_{\theta, \epsilon}}
         K({p} \| {p}_{\theta}) \bigr]\\
     &=& -\varliminf_{\epsilon \to 0} \frac{1}{\epsilon^2}
      \inf_{\theta^{\prime} \in \Theta^{o}}\{ 
        K({p}_{\theta^{\prime}}\| {p}_{\theta});  \Delta (\theta^{\prime}, \theta) >\epsilon^2 \}\\
     &=& - \frac{1}{\sigma_{1}(\sqrt{H_{\theta}}F^{-1}_{\theta}\sqrt{H_{\theta}})},\label{eq:suplim_explicit_form}
   \end{eqnarray}
   where we used Lemma 3 in the last line.
   Because a maximum likelihood estimator satisfies both
   Eqs.(\ref{eq:Amain1}) and (\ref{eq:suplim_explicit_form}), it
   achieves the equality in Eq.(\ref{eq:Amain1}), and Eq.(\ref{eq:lemma5}) holds.
   $\square$

% Lemma 6   
   The final lemma relates the identifiability condition in
   classical statistical estimation theory to informational
   completeness in quantum tomography. 
   We assume now that the probability distributions are given by 
   quantum mechanics, Eq.(\ref{eq:p=rhopi}), for finite dimensional systems.
   \begin{lemma}\label{lemma:6}
      Let $\hat{\rho} = \hat{\rho}(\bm{s})$ denote a density operator parametrized by a
     vector $\bm{s} \in S$.
      We assume that the parametrization is one-to-one.
      Suppose that we perform quantum state tomography with a POVM 
      $\bm{\Pi}=\{ \hat{\Pi}_{x} \}_{x \in \Omega}$.
      Then the following statements are equivalent.
      \begin{enumerate}
      \item The probability distribution describing the tomographic experiment satisfies the identifiability
       condition.\label{cnd:1}
   
      \item The Fisher matrix $F_{\bm{s}}$ is full rank for any $\bm{s} \in
       S^{o}$.\label{cnd:2}
   
      \item The POVM is informationally complete.\label{cnd:3}
      \end{enumerate}
   \end{lemma}
    {\bf Proof:}
% Linear parameterization is sufficient.
      First we show that it is sufficient to prove the equivalence of the
    three conditions in Lemma \ref{lemma:6} for a linear parametrization.
    In quantum mechanics, for a finite dimensional system, any
    probability distribution is linearly one-to-one parametrizable, and
    we can assume that the probability distribution has the form
    \begin{eqnarray}
      p_{\bm{s}}(x ) = v(x) + \bm{s}\cdot \bm{w}(x),\label{eq:Lparametrization}
    \end{eqnarray} 
    where $v(x)\in \mathbf{R}$ and $\bm{w}(x) \in \mathbf{R}^{d^2 -1}$ satisfy
    $\sum_{x\in \Omega}p_{\bm{s}}(x)=1$ for any $\bm{s} \in S$.
    If the probability distribution is one-to-one (but not necessarily linearly) parametrized by a different
    parameter $\bm{t}\in \mathbf{R}^{d^2 -1}$, then we have
    \begin{eqnarray}
      \tilde{p}_{\bm{t}}(x ) = p_{\bm{s}(\bm{t})}(x ),\\
      \tilde{\nabla}_{\bm{t}}\tilde{p}_{\bm{t}}(x ) = \frac{\partial
     \bm{s}}{\partial{\bm{t}}} \nabla_{\bm{s}}p_{\bm{s}}(x ).
    \end{eqnarray}
    Condition \ref{cnd:1} for $\bm{s}$ and condition \ref{cnd:1} for
    $\bm{t}$ are equivalent because both parametrizations are one-to-one.
    Condition \ref{cnd:2} for $\bm{s}$ and condition \ref{cnd:2} for
    $\bm{t}$ are equivalent because the Fisher matrices satisfy the equation
    \begin{eqnarray}
       \tilde{F}_{\bm{t}} = \frac{\partial
     \bm{s}}{\partial{\bm{t}}} F_{\bm{s}}\frac{\partial
     \bm{s}}{\partial{\bm{t}}}^{T},
    \end{eqnarray}  
    and the Jacobian $\frac{\partial \bm{s}}{\partial{\bm{t}}}$ is full
    rank.
    Condition \ref{cnd:3} is independent of state parametrization.
    Therefore if condition \ref{cnd:1}, \ref{cnd:2}, and \ref{cnd:3} are
    equivalent for a linear parametrization, then they are also equivalent
    for a general parametrization. 

% Equivalence between condition 1 and condition 2
    Next we prove the equivalence of conditions \ref{cnd:1} and \ref{cnd:2}.    
    As in the above discussion, without loss of generality, we can
    assume that $\bm{s}$ is the fixed parameter such
    that ${F}_{\bm{s}}$ is diagonalized because this is a linear transformation of a general parameter.
    Under this assumption, condition \ref{cnd:1} is equivalent to the
    condition that for any $\bm{s} \in S^{o}$ and for all $\alpha = 1, \ldots , d^2 -1$,
    there exists at least one $x \in \Omega$ such that
    \begin{eqnarray}
      \partial_{\alpha}p_{\bm{s}}(x)\neq 0, \label{cnd:A}
    \end{eqnarray}
    where $\partial_{\alpha}:=\frac{\partial}{\partial s^{\alpha}}$.
    On the other hand, the diagonal elements of the Fisher matrix are
    \begin{eqnarray}
       F_{\bm{s},\alpha\alpha} = \sum_{x \in \Omega} \frac{(\partial_{\alpha}p_{\bm{s}}(x
     ))^2}{p_{\bm{s}}(x )},\ \alpha = 1, \ldots , d^2 -1.
    \end{eqnarray}
    Therefore the full rankness of the Fisher matrix is equivalent to
    Eq.(\ref{cnd:A}), and condition \ref{cnd:1} and condition
    \ref{cnd:2} are equivalent.

% Equivalence between condition 2 and condition 3 
    Third we prove the equivalence between condition \ref{cnd:2} and
    \ref{cnd:3}.
      We choose the generalized Bloch parametrization of density
    operators \cite{Kimura03, Byrd03}; any density operator $\hat{\rho}$ can be
    represented as
    \begin{eqnarray}
       \hat{\rho}(\bm{s})= \frac{1}{d}\hat{I}+\frac{1}{2}\bm{s}\cdot \hat{\bm{\sigma}},
    \end{eqnarray}
    where $\hat{I}$ is the identity operator on $\mathcal{H}$ and
    $\hat{\sigma}_{\alpha}$ are generators of $SU(d)$ satisfying
    $\hat{\sigma}_{\alpha}=\hat{\sigma}_{\alpha}^{\dagger}$,
    $\mbox{Tr}[\hat{\sigma}_{\alpha}]=0$,
    and
    $\mbox{Tr}[\hat{\sigma}_{\alpha}\hat{\sigma}_{\beta}]=2\delta_{\alpha
    , \beta}$ ($\alpha, \beta = 1, \ldots , d^2 -1$).
    To determine the representation uniquely, we need more additional
    conditions on $\hat{\bm{\sigma}}$, but the additional conditions are
    not used in the following discussion.
    Each element of the tester POVM can be represented as
    \begin{eqnarray}     
       \hat{\Pi}_{x} = v(x)\hat{I}+ \bm{w}(x)\cdot \hat{\bm{\sigma}},\
	x\in \Omega ,  
    \end{eqnarray}
    where $v(x)$ and $\bm{w}(x)$ should satisfy $\sum_{x \in \Omega}v(x)=1$, 
      $\sum_{x\in \Omega}\bm{w}(x)=\bm{0}$, and $\hat{\Pi}_{x}\ge 0$ for
      any $x\in \Omega$.
    Then, the probability distribution describing the tomographic
      experiment is represented as Eq.(\ref{eq:Lparametrization}),
%      \begin{eqnarray}
%         p_{\bm{s}}(x)&=\mbox{Tr}[\hat{\rho}(\bm{s})\hat{\Pi}_{x}]\\
%         &= v_{x}+ \bm{w}_{x}\cdot \bm{s},
%      \end{eqnarray} 
      and the Fisher matrix is
      \begin{eqnarray}
         F_{\bm{s}}&=& \sum_{x \in \Omega} p_{\bm{s}}(x)\nabla_{\bm{s}}\log
       p_{\bm{s}}(x) \nabla_{\bm{s}}\log p_{\bm{s}}(x)^{T}\\
         &=& \sum_{x \in \Omega}\frac{\bm{w}(x)\bm{w}(x)^{T}}{p_{\bm{s}}(x)}.
      \end{eqnarray}
      Therefore the full rankness of the Fisher matrix is equivalent to the
    condition that $\{\bm{w}(x)\}_{x\in \Omega}$ spans
    $\mathbf{R}^{d^2 -1}$, and this implies that the tester POVM
    $\bm{\Pi}$ is informationally complete.   
      $\square$

\subsection{A more general theorem}

From Eq.(\ref{eq:Amain1}) and Eq.(\ref{eq:lemma5}), we obtain the following theorem.
% Theorem 2
   \renewcommand{\thetheorem}{2}
   \begin{theorem}\label{theorem:2}
     Suppose that $\Delta$ on $\Theta$ is a sufficiently smooth pseudo-distance with a non-zero same
    point Hesse matrix $H_{\theta}$.
     If $\theta \in \Theta^{o}$, for an arbitrary consistent estimator $\theta^{\mathrm{est}}$,
     the following inequality holds: 
     \begin{eqnarray}
      \varliminf_{\epsilon \to 0}\varliminf_{N\to \infty}\frac{1}{\epsilon^2 N}\log
     P^{(N)}_{\theta}(\Delta(\theta^{\mathrm{est}}_{N}, \theta)> \epsilon^2 ) \notag \\
     \ge -1/\sigma_{1}( \sqrt{H_{\theta}}F_{\theta}^{-1}\sqrt{H_{\theta}}).\label{eq:lemma1-1}
   % -\inf_{d(\theta , \theta)\ge \epsilon}\{ K(\theta | \theta) \}.
     \end{eqnarray}
     Furthermore, when the identifiability condition is satisfied,
     a maximum likelihood estimator $\theta^{\mathrm{ml}}$ is
    consistent and achieves the equality in Eq.(\ref{eq:lemma1-1}),
    {\it i.e.}, 
     \begin{eqnarray}
      \lim_{\epsilon \to 0}\lim_{N\to \infty}\frac{1}{\epsilon^2 N}\log
     P^{(N)}_{\theta}(\Delta(\theta^{\mathrm{ml}}_{N}, \theta)> \epsilon^2 ) \notag \\
     = -1/\sigma_{1}( \sqrt{H_{\theta}}F_{\theta}^{-1}\sqrt{H_{\theta}})\label{eq:lemma1-2}
   % -\inf_{d(\theta , \theta)\ge \epsilon}\{ K(\theta | \theta) \}.
     \end{eqnarray}
    holds. 
   \end{theorem}
   Theorem 2 is in fact more general than Theorem 1, since identifiability
   is more general than informational completeness.
   Hence, the properties that the error probabilities of consistent
   estimators can decrease at most exponentially, the rate of decrease is bounded
   by the maximal eigenvalue of a matrix, and the bound is achievable by
   a maximum-likelihood estimator are common to a larger class of
   probability distributions than those of quantum mechanics.

   By applying Theorem 2 to quantum state tomography and using Lemma 6, we can obtain Theorem 1.
   Theorem 2 is applicable to the other types of quantum tomography.
   The conditions corresponding to the identifiability condition are
   different, and can be derived in the same way as in the proof of Lemma 6.
   For example, let us consider ancilla-unassisted quantum process tomography.
   To identify an unknown quantum process
   described by a linear, completely-positive, and trace-preserving map
   $\kappa$ on $\mathcal{S}(\mathcal{H})$,
   we prepare a set of input states $\bm{\rho}=\{ \hat{\rho}_{n} \}_{n=1}^{N_s}$
   where $\hat{\rho}_{n} \in \mathcal{S}(\mathcal{H})$ and a
   measurement described by a POVM $\bm{\Pi}=\{\hat{\Pi}_{x}\}_{x\in \Omega}$ on $\mathcal{H}$.
   The set $\{ \bm{\rho}, \bm{\Pi} \}$ is the tester for ancilla-unassisted process tomography.
   When $\bm{\rho}$ spans $\mathcal{S}(\mathcal{H})$, it is called tomographically complete.
   In ancilla-unassisted process tomography, the informational completeness of $\bm{\Pi}$ 
   and the tomographical completeness of $\bm{\rho}$ both are required.
   We can prove that these conditions are equivalent to the
   identifiability condition in the same way as in lemma \ref{lemma:6}.

% A generalized identifiability condition
   For the case where the same point Hesse matrix of the loss function
   is positive semidefinite, as mentioned in subsection \ref{subsection:general}, 
   the identifiability condition is modified as follows:
   for any $\theta \in \Theta^{o}$ and $\theta^{\prime} \in \Theta$, if $g(\theta)\neq
    g(\theta^{\prime})$, then there exists at least one single outcome
    $x \in \Omega$ satisfying $p_{\theta}(x) \neq p_{\theta^{\prime}}(x)$.
   Theorem 2 holds for this modification.

\begin{acknowledgments}
   T.S. thanks Yu Watanabe for helpful discussions.  
   P.S.T. acknowledges useful discussion with P. Kim.
   This work was supported by JSPS Research Fellowships for Young Scientists, 
   JSPS KAKENHI (20549002) for Scientific Research (C),
   and Special Coordination Funds for Promoting Science and Technology.
\end{acknowledgments}
%merlin.mbs 2010-03-15 4.21a (PWD, AO, DPC)
\end{document}